\documentclass{ifacconf}

\usepackage{graphicx}      
\usepackage{natbib}        

\expandafter\let\csname theoremstyle\endcsname\relax

\usepackage{amsmath,amssymb,amsthm}
\usepackage{graphicx}
\usepackage{booktabs}
\usepackage{color}
\usepackage{multicol}
\usepackage{makecell}
\usepackage{balance}
\usepackage{subcaption}

\theoremstyle{theorem}
\newtheorem{theorem}{Theorem}
\theoremstyle{definition}

\newtheorem{lemma}{Lemma}
\theoremstyle{remark}
\newtheorem{remark}{Remark}

\begin{document}
\begin{frontmatter}

\title{A Unidirectionally Connected FAS Approach for 6-DOF Quadrotor Control
\thanksref{footnoteinfo}} 

\thanks[footnoteinfo]{
Corresponding author: Guang-Ren Duan.
This work was supported in part by the Science Center Program of the National Natural Science Foundation of China (NSFC) under Grant 62188101 and NSFC under Grant 623B2045, in part by the Guangdong Provincial Natural Science Foundation under Grant 2024A1515011648, and in part by the Shenzhen Science and Technology Program under Grant KQTD20221101093557010.}

\author[First]{Weijie Ren} 
\author[First]{Haowen Liu} 
\author[Second]{Guang-Ren Duan}

\address[First]{Guangdong Provincial Key Laboratory of Fully Actuated System Control Theory and Technology, Southern University of Science and Technology, Shenzhen 518055, China (e-mail: weijie.ren@outlook.com, 12132633@mail.sustech.edu.cn).}
\address[Second]{Guangdong Provincial Key Laboratory of Fully Actuated System Control Theory and Technology, Southern University of Science and Technology, Shenzhen 518055, China, and the Center for Control Theory and Guidance Technology, Harbin Institute of Technology, Harbin 150001, China (e-mail: g.r.duan@hit.edu.cn)}

\begin{abstract}                
This paper proposes a unidirectionally connected fully actuated system (UC-FAS) approach for the sub-stabilization and tracking control of 6-DOF quadrotors, tackling limitations both in state-space and FAS framework to some extent. The framework systematically converts underactuated quadrotor dynamics into a UC-FAS model, unifying the existing different FAS transformation ways. By eliminating estimation of the high-order derivatives of control inputs, a drawback of current methods, the UC-FAS model simplifies controller design and enables direct eigenstructure assignment for closed-loop dynamics. Simulations demonstrate precise 6-DOF tracking performance. This work bridges theoretical FAS approach advancements with practical implementation needs, offering a standardized paradigm for nonlinear quadrotor control.
\end{abstract}

\begin{keyword}
Fully actuated system approach, nonlinear control, quadrotor, sub-stabilization, trajectory tracking, unidirectionally connected system
\end{keyword}

\end{frontmatter}

\section{Introduction}


The quadrotor, a type of unmanned aerial vehicle (UAV), is valued for its simple design, maneuverability, and applications like aerial photography, surveillance, and autonomous delivery \cite{Khalid2023ACMCS_quadrotor}. It uses four rotors to generate lift and control motion, allowing for agile movement in complex environments.
However, controlling quadrotors is challenging due to their underactuated nature, nonlinear dynamics, and sensitivity to external disturbances \cite{Emran2018ARC}. The coupling between translational and rotational dynamics requires advanced control strategies, and actuator faults further highlight the need for robust, fault-tolerant control \cite{Ke2023TRO_FTC}.

Various control techniques have been proposed to address quadrotor challenges, from classical PID controllers \cite{Lopez2023ARC} to modern methods like feedback linearization (FL) \cite{Choi2015TMech,Lotufo2020TCST_Embedded}, model predictive control (MPC) \cite{Nan2022RAL_MPC}, and reinforcement learning \cite{Hua2023TIE}. Nonlinear control approaches offer significant advantages over local linearization models. \cite{Choi2015TMech} introduced a backstepping-like FL method for quadrotor tracking, assuming zero derivatives of desired position and angular rates, limiting its ability to handle time-varying trajectories. Additionally, \cite{Choi2015TMech} noted that control parameters must be carefully chosen for stability, but did not provide a thorough analysis of the expected closed-loop system, a common issue in FL-based approaches \cite{Lotufo2020TCST_Embedded}. Furthermore, applying standard FL to underactuated systems like quadrotors remains challenging due to complex design procedures \cite{Emran2018ARC}.

The existing methods, totally based on the state-space framework, often lead to complex controller design for underactuated quadrotors because they focus on state variables, rather than control inputs.

Recently, Duan introduced a new control framework called the fully actuated system (FAS) approach, which artificially transforms the original underactuated system model into a FAS one in the sense of mathematics \cite{Duan_IJSS_I_Models,Duan_IJSS_VII_Controllability}. The derived FAS model allows for straightforward controller design using full-actuation characteristic, canceling system dynamics if desired, and providing a linear closed-loop system with assigned eigenstructure. The FAS approach has been successfully applied in systems like quadrotors, robot arms, and flexible servo systems \cite{Duan2024TCYB_FASoverview,RenDuanLiKong2025TMech}.

The FAS approach for quadrotors has gained attention \cite{Lu2022CSL_3-DOF_quadrotor,Lu2023ACC_6-DOF_Quadrotor,Lu2024FASTA,Wang2024ISATrans_quadrotor,Xu2024TAES_FTC_Quadrotor}, such as the work on 3-DOF and 6-DOF quadrotor tracking control in \cite{Lu2022CSL_3-DOF_quadrotor,Lu2023ACC_6-DOF_Quadrotor}. The main contribution is that these two papers proposed a novel trick to convert the quadrotor model into a mix-order FAS whereas the standard method in \cite{Duan_IJSS_I_Models} is proven to be failed. Based on a more comprehensive framework, \cite{Wang2024ISATrans_quadrotor} investigated the model construction and predictive control for quadrotor UAV information gathering tracking missions. Moreover, fault-tolerant control has been studied in \cite{Lu2024FASTA,Xu2024TAES_FTC_Quadrotor}.

Despite the above achievements obtained in the field of quadrotor control using the FAS approach, a common problem is that different research provides varying methods for deriving FAS models from quadrotor dynamics. For example, \cite{Lu2022CSL_3-DOF_quadrotor,Lu2023ACC_6-DOF_Quadrotor} proved that the quadrotor model can not be transformed into a general FAS, since the unimodular transformation matrix does not exist. The authors thereby proposed a modified FAS approach. However, other papers also produced their FAS models (see \cite{Wang2024ISATrans_quadrotor,Xu2024TAES_FTC_Quadrotor}). This inconsistency, i.e., the appearance of different FAS models, may cause confusion. A reasonable explanation could be that the existing literature inspected the model from different perspectives. Still, they did not realize that the FAS model of quadrotors belongs to a more generalized FAS, which may even describe a broader class of physical systems than the pioneering work proposed in \cite{Duan_IJSS_I_Models} (with a total of ten papers in the series).

Furthermore, the existing FAS approach for quadrotor control needs information on the first- and second-order derivatives of the total rotor thrust (one of the input signals) \cite{Lu2022CSL_3-DOF_quadrotor,Lu2023ACC_6-DOF_Quadrotor,Lu2024FASTA,Wang2024ISATrans_quadrotor}, which is difficult or even unable to obtain directly, especially in experiments and applications. This challenge is common across state-space methods and other high-order approaches \cite{Lotufo2020TCST_Embedded}, requiring additional equipment or algorithms like extended state observers \cite{Lu2024FASTA}. Although the input derivatives can be estimated to some extent, problems, like the high computational burden and deterioration of control performance, will consequentially be introduced.

To handle the abovementioned gaps, in this paper, we realize the quadrotor control by introducing a more generalized FAS model called unidirectionally connected FAS (UC-FAS), which is originally proposed by \cite{Duan2025IJSS_UC-FAS_I,Duan2025IJSS_UC-FAS_II,Duan2025IJSS_UC-FAS_III} and can systematically describe the quadrotor system. Using the proposed UC-FAS approach, we achieve sub-stabilization and tracking control, completely overcoming the need for estimating input signal derivatives in our control framework. The contributions of this paper are summarized as follows:
\begin{enumerate}
	\item This paper introduces the UC-FAS approach for quadrotor sub-stabilization control for the first time. Unlike state-space frameworks in \cite{Emran2018ARC,Lopez2023ARC,Choi2015TMech,Lotufo2020TCST_Embedded}, the UC-FAS model provides new insight for nonlinear controller design, yielding a linear constant closed-loop system suitable for eigenstructure assignment. Compared to existing FAS approaches in \cite{Lu2022CSL_3-DOF_quadrotor,Lu2023ACC_6-DOF_Quadrotor,Lu2024FASTA,Wang2024ISATrans_quadrotor,Xu2024TAES_FTC_Quadrotor}, the proposed UC-FAS model is simpler and more standardized, eliminating the need for complex transformations.
	
	\item The difficulty in obtaining the first- and second-order derivatives of control inputs, both in the state-space \cite{Lotufo2020TCST_Embedded} and FAS frameworks \cite{Lu2022CSL_3-DOF_quadrotor,Lu2023ACC_6-DOF_Quadrotor,Lu2024FASTA,Wang2024ISATrans_quadrotor}, are completely solved in this paper using the UC-FAS approach and the control issues related to quadrotors are systematically analyzed.
\end{enumerate}

In this paper, we denote $\mathbb{R}^n$ as the $n$-dimensional Euclidean space. Let $A\in \mathbb{R}^{n\times m}$ represent a real matrix of dimensions $n$ by $m$. For any matrix $A$, $A^{T}$ indicates its transpose, while for any square matrix $B$, $B^{-1}$ represents its inverse. For a collection of matrices $A_i\in \mathbb{R}^{n\times m}$ where $i = 0,1,\ldots,k$, we define $A_{0\sim k} = [A_0 ~A_1 ~\cdots ~A_k]$.
For any vector $x\in \mathbb{R}^{n}$, we use $x^{(i)}$ to represent its $i$-th derivative where $i$ is the positive integer number and define $x^{(0\sim m)}=[x^T ~\dot{x}^T ~\cdots ~(x^{(m)})^T]^T$.

\section{System model and preliminaries}


The 6-DOF quadrotor model can be described as follows:
\begin{align}
	&
	\begin{cases}
		\ddot{x}=\frac{T}{m}\left(  \cos\phi\sin\theta\cos\psi+\sin\phi\sin\psi\right)
		\\
		\ddot{y}=\frac{T}{m}\left(  \cos\phi\sin\theta\sin\psi-\sin\phi\cos\psi\right)
		\\
		\ddot{z}=\frac{T}{m}\cos\phi\cos\theta-g,
	\end{cases}
	\label{a1}\\
	&
	\begin{cases}
		\dot{\phi}=p+q\sin\phi\tan\theta-r\cos\phi\tan\theta\\
		\dot{\theta}=q\cos\phi+r\sin\phi\\
		\dot{\psi}=-q\frac{\sin\phi}{\cos\theta}+r\frac{\cos\phi}{\cos\theta},%
	\end{cases}
	\label{a2}\\
	&
	\begin{cases}
		\dot{p}=\frac{J_{y}-J_{z}}{J_{x}}qr+\frac{\tau_{\phi}}{J_{x}}\\
		\dot{q}=\frac{J_{z}-J_{x}}{J_{y}}pr+\frac{\tau_{\theta}}{J_{y}}\\
		\dot{r}=\frac{J_{x}-J_{y}}{J_{z}}pq+\frac{\tau_{\psi}}{J_{z}},%
	\end{cases}
	\label{a3}%
\end{align}
where $x$, $y$, $z$ are the corresponding positions in the 3-D space, $\phi$, $\theta$, $\psi$ denote the roll, pitch, and yaw angles, respectively, and variables $p$, $q$, and $r$ denote the angular velocities of the quadrotor about the body-fixed $x$-, $y$-, and $z$-axes, respectively.
The moment of inertial in the $x$-, $y$-, and $z$-axis are represented by $J_x$, $J_y$, and $J_z$, respectively. There are four control inputs of the quadrotor system: $T$ representing the total rotor thrust and $\tau_{\phi}$, $\tau_{\theta}$, and $\tau_{\psi}$ being the body torques.
The angles have the subsequently described constraints:
\begin{align}
	\phi &  \in\Omega_{\phi}=\left\{  \phi|-\frac{\pi}{2}<\phi<\frac{\pi}%
	{2}\right\}  ,\label{b1}\\
	\theta &  \in\Omega_{\theta}=\left\{  \theta|-\frac{\pi}{2}<\theta<\frac{\pi
	}{2}\right\}  . \label{b2}
\end{align}
As a preparation, we define that
\[
\varPhi =\left[ \begin{matrix}
	\phi&		\theta&		\psi\\
\end{matrix} \right] ^T,~
\varLambda =\left[ \begin{matrix}
	p&		q&		r\\
\end{matrix} \right] ^T.
\]

The following two subsections are devoted to the technical handling of model equations (\ref{a2})-(\ref{a3}) and (\ref{a1}), respectively, as a preparation for proposing the UC-FAS model and the controller design.

\subsection{Treating dynamic equations (\ref{a2}) and (\ref{a3})}    \label{Treating dynamic equations a2 and a3}

Note that equations (\ref{a2}) and (\ref{a3}) are first-order ordinary differential equations that exhibit internal correlation with respect to input channels. Therefore, our goal is to transform them into a second-order equation.

Write (\ref{a2}) in the following form
\begin{equation}	\label{a5}
	\dot{\varPhi} = M\left(  \phi,\theta\right) \varLambda,
\end{equation}
where
$
M\left(  \phi,\theta\right)  =\left[
\begin{array}
	[c]{ccc}%
	1 & \sin\phi\tan\theta & -\cos\phi\tan\theta\\
	0 & \cos\phi & \sin\phi\\
	0 & -\frac{\sin\phi}{\cos\theta} & \frac{\cos\phi}{\cos\theta}%
\end{array}
\right]  .
$

Since $\det M\left(  \phi,\theta\right)  =-1/\cos\theta \neq0$ (due to the constraint (\ref{b2})), we can get from (\ref{a5}) that $\varLambda = M^{-1}\left(  \phi,\theta\right)  \dot{\varPhi}$.

Next, taking the differential of (\ref{a5}) and applying the relation in (\ref{a3}), yield
\begin{align}
	\begin{aligned}
		\ddot{\varPhi}&=\dot{M}\left( \phi ,\theta \right) \varLambda +M\left( \phi ,\theta \right) \dot{\varLambda}\\
		&=\dot{M}\left( \phi ,\theta \right) M^{-1}\left( \phi ,\theta \right) \dot{\varPhi}
		+M\left( \phi ,\theta \right) (\mathring{\varLambda} + \mathring{u}),
	\end{aligned}
\end{align}
where
$
\mathring{\varLambda}=\left[ \begin{array}{c}
	\frac{J_y-J_z}{J_x}qr\\
	\frac{J_z-J_x}{J_y}pr\\
	\frac{J_x-J_y}{J_z}pq\\
\end{array} \right] ,~
\mathring{u}=\left[ \begin{array}{c}
	\frac{\tau _{\phi}}{J_x}\\
	\frac{\tau _{\theta}}{J_y}\\
	\frac{\tau _{\psi}}{J_z}\\
\end{array} \right] .
$

Since $M\left(  \phi,\theta\right)  $ is nonsingular when constraint \eqref{b2} holds, we can introduce the following input transformation:
\begin{align}   \label{eq:input transformation of bu1bu2bu3}
	\bar{u}
	=\dot{M}\left( \phi ,\theta \right) M^{-1}\left( \phi ,\theta \right) \dot{\varPhi}
	+M\left( \phi ,\theta \right) (\mathring{\varLambda} + \mathring{u}).
\end{align}

Thus the system equations (\ref{a2}) and (\ref{a3}) become
$\ddot{\varPhi} = \bar{u}$,
where $\bar{u} = [\bar{u}_1 ~\bar{u}_2 ~\bar{u}_3]^T$. $\bar{u}_{1}$, $\bar{u}_{2}$, and $\bar{u}_{3}$ represent the intermediate input variables.

\subsection{Treating dynamic equation (\ref{a1})}    \label{Treating dynamic equations a1}

Noting (\ref{b1}) and (\ref{b2}), we can introduce
\begin{align}   \label{eq:input transformation of u0}
	u_{0}=\frac{T}{m}\cos\phi\cos\theta.
\end{align}
With the new input \eqref{eq:input transformation of u0} and system equation \eqref{a1}, we obtain
\begin{align*}
	&\ddot{x} = u_{0}\left(  \tan\theta\cos\psi+\tan\phi\sec\theta\sin\psi\right)  ,\\
	&\ddot{y} = u_{0}\left(  \tan\theta\sin\psi-\tan\phi\sec\theta\cos\psi\right)  .
\end{align*}
Therefore, the whole system can be arranged into
\begin{align}
	\begin{cases}
		\ddot{x}=u_{0}\left(  \tan\theta\cos\psi+\tan\phi\sec\theta\sin\psi\right) \\
		\ddot{y}=u_{0}\left(  \tan\theta\sin\psi-\tan\phi\sec\theta\cos\psi\right) \\
		\ddot{z}=u_{0}-g \\
        \ddot{\varPhi}=\bar{u}.
	\end{cases}
	\label{a6}
\end{align}

\begin{remark}
	Subsection \ref{Treating dynamic equations a2 and a3} employs an input transformation to simplify the design and also implies that $\theta$ should stay within $-\frac{\pi}{2}$ to $\frac{\pi}{2}$. Subsection \ref{Treating dynamic equations a1} plays a trick in defining a new input $u_0$ instead of using $T$, allowing us easily to obtain the analytical form of the first- and second-order derivatives of $u_0$ later. Since $u_0$ is entirely up to users to determine based on the UC-FAS approach, the extended state observer, for estimating $u_0^{(1\sim 2)}$ in \cite{Lu2024FASTA}, is unnecessary.
\end{remark}

\section{The constrained UC-FAS model}

We first denote
\begin{align*}
	f_{x}\left(  \phi,\theta,\psi\right)   &  =\tan\theta\cos\psi+\tan\phi
	\sec\theta\sin\psi,\\
	f_{y}\left(  \phi,\theta,\psi\right)   &  =\tan\theta\sin\psi-\tan\phi
	\sec\theta\cos\psi.
\end{align*}
Then the dynamic equations regarding $x$ and $y$ given in \eqref{a6} can be expressed by%
\[%
\begin{cases}
	\ddot{x}=u_{0}f_{x}\left(  \phi,\theta,\psi\right) \\
	\ddot{y}=u_{0}f_{y}\left(  \phi,\theta,\psi\right)  .
\end{cases}
\]

Taking the time derivative of $\ddot{x}$\ yields
\begin{align*}
	\dddot{x}=\dot{u}_0f_x\left( \phi ,\theta ,\psi \right) +u_0\Gamma _x\left( \phi ,\theta ,\psi \right) \dot{\varPhi},
\end{align*}
where
$
\Gamma _x\left( \phi ,\theta ,\psi \right) =\left[ \begin{matrix}
	\frac{\partial}{\partial \phi}f_x\left( \cdot \right)&		\frac{\partial}{\partial \theta}f_x\left( \cdot \right)&		\frac{\partial}{\partial \psi}f_x\left( \cdot \right)\\
\end{matrix} \right] .
$

Further, taking the time derivative of $\dddot{x}$ again produces%
\begin{align*}
	x^{\left( 4 \right)}=\ddot{u}_0f_x\left( \cdot \right) +\left[ 2\dot{u}_0\Gamma _x\left( \cdot \right) +u_0\dot{\Gamma}_x\left( \cdot \right) \right] \dot{\varPhi}+u_0\Gamma _x\left( \cdot \right) \bar{u}.
\end{align*}
where
\begin{align*}
	\Gamma _x\left( \phi ,\theta ,\psi \right) =\left[ \begin{matrix}
		\Gamma _{x,1}\left( \phi ,\theta ,\psi \right)&		\Gamma _{x,2}\left( \phi ,\theta ,\psi \right)&		\Gamma _{x,3}\left( \phi ,\theta ,\psi \right)\\
	\end{matrix} \right] ,
\end{align*}
\begin{align*}
	\Gamma_{x,1}\left(  \phi,\theta,\psi\right)   &  =\sec^{2}\phi\sec\theta
	\sin\psi,\\
	\Gamma_{x,2}\left(  \phi,\theta,\psi\right)   &  =\sec^{2}\theta\cos\psi
	+\tan\phi\sec\theta\tan\theta\sin\psi,\\
	\Gamma_{x,3}\left(  \phi,\theta,\psi\right)   &  =-\tan\theta\sin\psi+\tan
	\phi\sec\theta\cos\psi,
\end{align*}
and
$
\dot{\Gamma}_{x}\left(  \phi,\theta,\psi\right)  =\left[
\begin{array}
	[c]{ccc}%
	\dot{\Gamma}_{x,1} & \dot{\Gamma}_{x,2} & \dot{\Gamma}_{x,3}%
\end{array}
\right]  ,
$
where
\begin{align*}
	&\begin{aligned}
		\dot{\Gamma}_{x,1}
		=&2\dot{\phi}\sec\phi\tan\phi\sec\theta\sin\psi
		+\dot{\theta}\sec^{2}\phi\sec\theta\tan\theta\sin\psi\\
		&  +\dot{\psi}\sec^{2}\phi\sec\theta\cos\psi,
	\end{aligned}\\
	&\begin{aligned}
		\dot{\Gamma}_{x,2}
		=&2\dot{\theta}\sec\theta\tan\theta\cos\psi-\dot{\psi
		}\sec^{2}\theta\sin\psi\\
		&  +\dot{\phi}\sec^{2}\phi\sec\theta\tan\theta\sin\psi+\dot{\theta}\tan
		\phi\sec\theta\tan^{2}\theta\sin\psi\\
		&  +\dot{\theta}\tan\phi\sec^{3}\theta\sin\psi+\dot{\psi}\tan\phi\sec
		\theta\tan\theta\cos\psi,
	\end{aligned}\\
	&\begin{aligned}
		\dot{\Gamma}_{x,3}
		=&-\dot{\theta}\sec^{2}\theta\sin\psi-\dot{\psi}
		\tan\theta\cos\psi+\dot{\phi}\sec^{2}\phi\sec\theta\cos\psi\\
		&  +\dot{\theta}\tan\phi\sec\theta\tan\theta\cos\psi-\dot{\psi}\tan\phi
		\sec\theta\sin\psi.
	\end{aligned}
\end{align*}

The similar process can be realized on $\ddot{y}$, so that we obtain
\begin{align*}
	y^{\left( 4 \right)}=\ddot{u}_0f_y\left( \cdot \right) +\left[ 2\dot{u}_0\Gamma _y\left( \cdot \right) +u_0\dot{\Gamma}_y\left( \cdot \right) \right] \dot{\varPhi}+u_0\Gamma _y\left( \cdot \right) \bar{u},
\end{align*}
where 
$
\Gamma _y\left( \phi ,\theta ,\psi \right) = \left[ \begin{matrix}
	\frac{\partial}{\partial \phi}f_y\left( \cdot \right)&		\frac{\partial}{\partial \theta}f_y\left( \cdot \right)&		\frac{\partial}{\partial \psi}f_y\left( \cdot \right)\\
\end{matrix} \right]\\
=\left[ \begin{matrix}
	\Gamma _{y,1}\left( \phi ,\theta ,\psi \right)&		\Gamma _{y,2}\left( \phi ,\theta ,\psi \right)&		\Gamma _{y,3}\left( \phi ,\theta ,\psi \right)\\
\end{matrix} \right] ,
$
\begin{align*}
	\Gamma_{y,1}\left(  \phi,\theta,\psi\right)   &  =-\sec^{2}\phi\sec\theta
	\cos\psi,\\
	\Gamma_{y,2}\left(  \phi,\theta,\psi\right)   &  =\sec^{2}\theta\sin\psi
	-\tan\phi\sec\theta\tan\theta\cos\psi,\\
	\Gamma_{y,3}\left(  \phi,\theta,\psi\right)   &  =\tan\theta\cos\psi+\tan
	\phi\sec\theta\sin\psi,
\end{align*}
and
$
\dot{\Gamma}_{y}\left(  \phi,\theta,\psi\right)  =\left[
\begin{array}
	[c]{ccc}
	\dot{\Gamma}_{y,1} & \dot{\Gamma}_{y,2} & \dot{\Gamma}_{y,3}%
\end{array}
\right]  ,
$
where
\begin{align*}
	\dot{\Gamma}_{y,1}    =&-2\dot{\phi}\sec\phi\tan\phi\sec\theta\cos\psi
	-\dot{\theta}\sec^{2}\phi\sec\theta\tan\theta\cos\psi\\
	&  +\dot{\psi}\sec^{2}\phi\sec\theta\sin\psi,\\
	\dot{\Gamma}_{y,2}    =&2\dot{\theta}\sec\theta\tan\theta\sin\psi+\dot{\psi
	}\sec^{2}\theta\cos\psi\\
	&-\dot{\phi}\sec^{2}\phi\sec\theta\tan\theta\cos\psi-\dot{\theta}\tan\phi\sec\theta\tan^{2}\theta\cos\psi\\
	&-\dot{\theta}\tan\phi\sec^{3}\theta\cos\psi+\dot{\psi}\tan\phi\sec\theta\tan\theta\sin\psi,\\
	\dot{\Gamma}_{y,3}    =&\dot{\theta}\sec^{2}\theta\cos\psi-\dot{\psi}
	\tan\theta\sin\psi+\dot{\phi}\sec^{2}\phi\sec\theta\sin\psi\\
	&  +\dot{\theta}\tan\phi\sec\theta\tan\theta\sin\psi+\dot{\psi}\tan\phi
	\sec\theta\cos\psi.
\end{align*}

From now on, the high-order dynamic equations regarding $x$ and $y$ can be
formulated into
\[
\left\{
	\begin{aligned}
		x^{\left( 4 \right)}=&g_x\left( \phi ^{\left( 0\sim 1 \right)},\theta ^{\left( 0\sim 1 \right)},\psi ^{\left( 0\sim 1 \right)},u_{0}^{\left( 0\sim 2 \right)},\bar{u}_3,t \right) \\
		&+G_x\left( \phi ,\theta ,\psi ,u_0 \right) u_2\\
		y^{\left( 4 \right)}=&g_y\left( \phi ^{\left( 0\sim 1 \right)},\theta ^{\left( 0\sim 1 \right)},\psi ^{\left( 0\sim 1 \right)},u_{0}^{\left( 0\sim 2 \right)},\bar{u}_3,t \right) \\
		&+G_y\left( \phi ,\theta ,\psi ,u_0 \right) u_2 ,\\
	\end{aligned}
\right. 
\]
where $u_2=\left[ \begin{matrix}
	\bar{u}_1&		\bar{u}_2\\
\end{matrix} \right] ^T$, and 
\begin{align*}
	&g_x\left( \phi ^{\left( 0\sim 1 \right)},\theta ^{\left( 0\sim 1 \right)},\psi ^{\left( 0\sim 1 \right)},u_{0}^{\left( 0\sim 2 \right)},\bar{u}_3,t \right)\\
	&=\ddot{u}_0f_x\left( \cdot \right) +\left[ 2\dot{u}_0\Gamma _x\left( \cdot \right) +u_0\dot{\Gamma}_x\left( \cdot \right) \right] \dot{\varPhi}+u_0\Gamma _{x,3}\left( \cdot \right) \bar{u}_3,\\
	&g_y\left( \phi ^{\left( 0\sim 1 \right)},\theta ^{\left( 0\sim 1 \right)},\psi ^{\left( 0\sim 1 \right)},u_{0}^{\left( 0\sim 2 \right)},\bar{u}_3,t \right)\\
	&=\ddot{u}_0f_y\left( \cdot \right) +\left[ 2\dot{u}_0\Gamma _y\left( \cdot \right) +u_0\dot{\Gamma}_y\left( \cdot \right) \right] \dot{\varPhi}+u_0\Gamma _{y,3}\left( \cdot \right) \bar{u}_3,
\end{align*}
\begin{align*}
	G_{x}\left(  \phi,\theta,\psi,u_{0}\right)   &  =u_{0}\left[
	\begin{array}
		[c]{cc}%
		\Gamma_{x,1}\left(  \phi,\theta,\psi\right)  & \Gamma_{x,2}\left(  \phi
		,\theta,\psi\right)
	\end{array}
	\right]  ,\\
	G_{y}\left(  \phi,\theta,\psi,u_{0}\right)   &  =u_{0}\left[
	\begin{array}
		[c]{cc}%
		\Gamma_{y,1}\left(  \phi,\theta,\psi\right)  & \Gamma_{y,2}\left(  \phi
		,\theta,\psi\right)
	\end{array}
	\right]  .
\end{align*}

By defining the new state $X=\left[ \begin{matrix}
	x&		y\\
\end{matrix} \right] ^T$ and $u_{1}=\bar{u}_{3}$,
we get the following described UC-FAS model:
\begin{align}
	&  \ddot{z}=u_{0}-g,\label{eq:UC-FAS_0}\\
	&  \ddot{\psi}=u_{1},\label{eq:UC-FAS_1}\\
	&  \begin{aligned}
		X^{\left(  4\right)  }=&g_{X}\left(  \phi^{\left(  0\sim1\right)
		},\theta^{\left(  0\sim1\right)  },\psi^{\left(  0\sim1\right)  }
		,u_{0}^{\left(  0\sim2\right)  },u_{1},t\right)  \\
		&+G_{X}\left(  \phi
		,\theta,\psi,u_{0}\right)  u_{2},
	\end{aligned} \label{eq:UC-FAS_2}
\end{align}

where
\[
\begin{array}{l}
	g_{X}\left( \phi ^{\left( 0\sim 1 \right)},\theta ^{\left( 0\sim 1 \right)},\psi ^{\left( 0\sim 1 \right)},u_{0}^{\left( 0\sim 2 \right)},u_1,t \right)\\
	=\left[ \begin{array}{c}
		g_x\left( \phi ^{\left( 0\sim 1 \right)},\theta ^{\left( 0\sim 1 \right)},\psi ^{\left( 0\sim 1 \right)},u_{0}^{\left( 0\sim 2 \right)},u_1,t \right)\\
		g_y\left( \phi ^{\left( 0\sim 1 \right)},\theta ^{\left( 0\sim 1 \right)},\psi ^{\left( 0\sim 1 \right)},u_{0}^{\left( 0\sim 2 \right)},u_1,t \right)\\
	\end{array} \right],
\end{array}
\]
\[
G_{X}\left( \phi ,\theta ,\psi ,u_0 \right) =\left[ \begin{array}{c}
	G_x\left( \phi ,\theta ,\psi ,u_0 \right)\\
	G_y\left( \phi ,\theta ,\psi ,u_0 \right)\\
\end{array} \right],
\]

Next, the nonsingularity of matrix $G_{X}\left(  \phi,\theta,\psi,u_{0}\right)  $ is analyzed as follows:
\begin{align*}
    G_{X}\left( \phi ,\theta ,\psi ,u_0 \right) =\left[ \begin{array}{c}
        G_x\left( \cdot \right)\\
        G_y\left( \cdot \right)\\
    \end{array} \right] =u_0\left[ \begin{matrix}
        \Gamma _{x,1}\left( \cdot \right)&		\Gamma _{x,2}\left( \cdot \right)\\
        \Gamma _{y,1}\left( \cdot \right)&		\Gamma _{y,2}\left( \cdot \right)\\
    \end{matrix} \right].
\end{align*}
Calculate the determinant of $G_{X}\left(  \phi,\theta,\psi,u_{0}\right)  $
yielding
\[
\det\left(  G_{X}\left(  \phi,\theta,\psi,u_{0}\right)  \right)  =\frac{u_{0}}{\cos^{3}\theta\cos^{2}\phi}.
\]
Thus, if $\left\vert \theta\right\vert \leq\frac{\pi}{2}$, $\left\vert
\phi\right\vert \leq\frac{\pi}{2}$, i.e., constraints \eqref{b1} and \eqref{b2}, and the control input $u_{0}\neq0$, then
$\det G_{X}\left(  \cdot \right)  \neq0$, which means
$G_{X}\left(  \cdot \right)  $ is nonsingular, so that the full-actuation is achieved.
Meanwhile, considering the practical hardware implementation, the inputs are constrained by
\begin{align}
	u_{0}^{\left( 0\sim 2 \right)} & \in \mathcal{U} _0=\left\{ u_{0}^{\left( 0\sim 2 \right)} \middle| \begin{array}{c}
		u_0 \neq 0, ~\underline{u}_0\le u_0\le \bar{u}_0,\\
		\underline{u}_{0}^{\mathrm{d}1}\le \dot{u}_0\le \bar{u}_{0}^{\mathrm{d}1},\\
		\underline{u}_{0}^{\mathrm{d}2}\le \ddot{u}_0\le \bar{u}_{0}^{\mathrm{d}2}
	\end{array} \right\}   ,\label{eq:constraint u0}\\
	u_{1}  &  \in\mathcal{U}_{1}=\left\{  u_{1}|\underline{u}_{1}\leq u_{1}
	\leq\bar{u}_{1}\right\}  ,\label{eq:constraint u1}\\
	u_{2}  &  \in\mathcal{U}_{2}=\left\{  u_{2}|\underline{u}_{2}\leq u_{2}
	\leq\bar{u}_{2}\right\}  , \label{eq:constraint u2}
\end{align}
where $\underline{u}_0, \bar{u}_0, \underline{u}_{0}^{\mathrm{d}1}, \bar{u}_{0}^{\mathrm{d}1}, \underline{u}_{0}^{\mathrm{d}2}, \bar{u}_{0}^{\mathrm{d}2}, \underline{u}_{1}, \bar{u}_{1}, \underline{u}_{2}$, and $\bar{u}_{2}$ are corresponding constant upper and lower bounds.
Consequently, the following theorem about UC-FAS model is proposed.
\begin{theorem}
    System \eqref{eq:UC-FAS_0}-\eqref{eq:UC-FAS_2} is the UC-FAS model for the quadrotor dynamics \eqref{a1}-\eqref{a3}, subject to the constraints \eqref{b1}, \eqref{b2}, and \eqref{eq:constraint u0}-\eqref{eq:constraint u2}.
\end{theorem}

\begin{remark}
	As a supplementary note, we observe that in addition to states, the previous inputs $u_0$, $u_1$, and the corresponding time derivatives appear in \eqref{eq:UC-FAS_2}, which construct a form of UC-FAS. For further details, refer to \cite{Duan2025IJSS_UC-FAS_I,Duan2025IJSS_UC-FAS_II,Duan2025IJSS_UC-FAS_III}.
\end{remark}

\section{Sub-stabilization controller design}   \label{sec:Sub-stabilization controller design}

In this section, we design the sub-stabilization controllers for the 
constrained UC-FAS (\ref{eq:UC-FAS_0})-(\ref{eq:UC-FAS_2}) step by step.

\subsection{Sub-stabilizing subsystem (\ref{eq:UC-FAS_0})}

Following the subsystem equation (\ref{eq:UC-FAS_0}) of the derived
constrained UC-FAS, the corresponding internal feasibility set is easily
identified as
$
\mathbb{F}_{z}^{\mathrm{in}}=\left\{  Z|Z\in
\mathbb{R}^{2}\right\},
$
where we denote $Z = z^{(0 \sim 1)}$.
Then, the sub-stabilization controller can be proposed as
\begin{equation}
	u_{0}=-A_{0,0\sim1}z^{\left(  0\sim1\right)  }+g,
	\label{eq:stabilization controller_0}
\end{equation}
where $A_{0,0\sim1}$ is the parameter matrix to be designed, and the closed-loop is obtained
by
\begin{equation}	\label{eq:closed-loop system_0}
	\ddot{z}+A_{0,0\sim1}z^{\left(  0\sim1\right)  }=0,
\end{equation}
which is equivalent to
$
\dot{z}^{\left(  0\sim1\right)  }=\Upsilon\left(  A_{0,0\sim1}\right)  z^{\left(0\sim1\right)  }
$,
and $\Upsilon\left(  A_{0,0\sim1}\right)$ is the state transfer matrix derived from \eqref{eq:closed-loop system_0}.
Hence, the state response of the closed-loop system
(\ref{eq:closed-loop system_0}) can be given by
$
Z\left(  t\right)  =e^{\Upsilon\left(  A_{0,0\sim1}\right)  t}Z_{0},
$
where $Z_{0}=Z\left(  0\right)  $ represents the initial condition of state
$z^{(0 \sim 1)}$.

Based on constraint (\ref{eq:constraint u0}), the external set of feasibility
is given by
$
\mathbb{F}_{z}^{\mathrm{ex}}=\left\{  Z|u_{0}^{\left(  0\sim2\right)  }\left(
Z,t\right)  \in\mathcal{U}_{0},
Z\in \mathbb{R}^{2}, t\geq0\right\}  .
$
The overall set of feasibility of subsystem (\ref{eq:UC-FAS_0}) subject to constraint
$
\mathbb{F}_{z}=\mathbb{F}_{z}^{\mathrm{in}}\cap\mathbb{F}_{z}^{\mathrm{ex}}.
$
Further, we define the following internal and external region of exponential attractions (RoEAs) of the subsystem
(\ref{eq:UC-FAS_0}), respectively, as
\begin{align*}
	\mathcal{R}_{z}^{\mathrm{in}}  &  =\left\{  Z_{0}|e^{\Upsilon\left(  A_{0,0\sim
			1}\right)  t}Z_{0}\in\mathbb{F}_{z}^{\mathrm{in}},t\geq0\right\}  ,\\
	\mathcal{R}_{z}^{\mathrm{ex}}  &  =\left\{  Z_{0}|e^{\Upsilon\left(  A_{0,0\sim
			1}\right)  t}Z_{0}\in\mathbb{F}_{z}^{\mathrm{ex}},t\geq0\right\}  ,
\end{align*}
and introduce the overall RoEA of the system as
\[
\mathcal{R}_{z}=\mathcal{R}_{z}^{\mathrm{in}}\cap\mathcal{R}_{z}^{\mathrm{ex}}.
\]
As a result, provided that the initial value $Z_{0}$ is selected within $\mathcal{R}_{z}$, the feasible response $Z(t)$, i.e., $z^{(0 \sim 1)}\left(  t\right)  $, is exponentially convergent within $\mathbb{F}_{z}$.

\subsection{Sub-stabilizing subsystem (\ref{eq:UC-FAS_1})}

Based on subsystem (\ref{eq:UC-FAS_1}), the internal feasibility set can be easily described by
$
\mathbb{F}_{\psi}^{\mathrm{in}}=\left\{  \Psi|\Psi\in
\mathbb{R}^{2}\right\},
$
with $\Psi = \psi^{(0 \sim 1)}$.
Then, the sub-stabilization controller can be proposed by
\begin{equation}	\label{eq:stabilization controller_1}
	u_{1}=-A_{1,0\sim1}\psi^{\left(  0\sim1\right)  },
\end{equation}
where $A_{1,0\sim1}$ is the parameter matrix to be designed, and the closed-loop system is obtained by
\begin{equation}	\label{eq:closed-loop system_1}
	\ddot{\psi}+A_{1,0\sim1}\psi^{\left(  0\sim1\right)  }=0,
\end{equation}
which is equivalent to
$
\dot{\psi}^{\left(  0\sim1\right)  }=\Upsilon\left(  A_{1,0\sim1}\right)
\psi^{\left(  0\sim1\right)  }.
$
Hence, the state response of the closed-loop system (\ref{eq:closed-loop system_1}) can be given by
$
\Psi(t)  =e^{\Upsilon\left(  A_{1,0\sim1}\right)  t}\Psi_{0},
$
where $\Psi_{0}=\Psi\left(  0\right)  $ represents the initial condition of
state $\psi^{(0 \sim 1)}$.

Considering constraint \eqref{eq:constraint u1}, the external set of feasibility is given by
$
\mathbb{F}_{\psi}^{\mathrm{ex}}=\left\{  \Psi|u_{1}\left(  \Psi,t\right)
\in\mathcal{U}_{1},\Psi\in\mathbb{R}^{2},t\geq0\right\}  .
$
The overall set of feasibility of subsystem (\ref{eq:UC-FAS_1}) is
$
\mathbb{F}_{\psi}=\mathbb{F}_{\psi}^{\mathrm{in}}\cap\mathbb{F}_{\psi
}^{\mathrm{ex}}.
$
Further, we define the overall RoEA of subsystem (\ref{eq:UC-FAS_1}) as
\[
\mathcal{R}_{\psi}=\left\{  \Psi_{0}|e^{\Upsilon\left(  A_{1,0\sim1}\right)
	t}\Psi_{0}\in\mathbb{F}_{\psi},t\geq0\right\}  .
\]
Then, provided that the initial value $\Psi_{0}$ is selected within $\mathcal{R}_{\psi}$, the feasible response $\Psi(t)$, i.e., $\psi^{(0 \sim 1)}(t)$, is exponentially convergent within $\mathbb{F}_{\psi}$.

\subsection{Sub-stabilizing subsystem (\ref{eq:UC-FAS_2})}

Finally, we come to handle the subsystem (\ref{eq:UC-FAS_2}), whose internal
feasibility set can be identified as
\[
\mathbb{F}_{X}^{\mathrm{in}}=\left\{  \textbf{X}|\det G_{X}\left(
\phi,\theta,\psi,u_{0}\right)  \neq0\text{ or }\infty,\textbf{X}\in\mathbb{R}^{8}\right\}  ,
\]
where $\textbf{X} = X^{(0 \sim 3)}$, or equivalently
\[
\mathbb{F}_{X}^{\mathrm{in}}=\left\{  \textbf{X}|\phi\in\Omega_{\phi
},\theta\in\Omega_{\theta},u_{0}\neq0,\textbf{X}\in\mathbb{R}^{8}\right\}  ,
\]
based on the singularity analysis of $G_{X}$.
Then, the sub-stabilization controller for (\ref{eq:UC-FAS_2}) can be designed as
\begin{equation}
	u_{2}=-G_{X}^{-1}\left(  \phi,\theta,\psi,u_{0}\right) \left( A_2 X^{\left(  0\sim3\right)  }+g_{X}\left( \cdot \right)  \right)  ,
	\label{eq:stabilization controller_2}
\end{equation}
where $A_2 = \mathrm{blkdiag} ( A_{2,0\sim3}^x, A_{2,0\sim3}^y )$ is the parameter to be determined,
and the closed-loop system is obtained by
\begin{equation}
	X^{\left(  4\right)  }+A_{2}X^{\left(  0\sim3\right)
	}=0, \label{eq:closed-loop system_2}
\end{equation}
which is equivalent to
$
\dot{X}^{\left(  0\sim3\right)  }=\mathrm{blkdiag} ( \Upsilon_x, \Upsilon_y )
X^{\left(  0\sim3\right)  },
$
where $\Upsilon_x = \Upsilon(  A_{2,0\sim3}^x )$, $\Upsilon_y = \Upsilon(  A_{2,0\sim3}^y )$,
and $A_{2,0\sim3}^x$, $A_{2,0\sim3}^y$ are the parameter matrices. Hence, the state response of the
closed-loop system (\ref{eq:closed-loop system_2}) can be given by
$
\textbf{X}\left(  t\right)  =e^{\mathrm{blkdiag} ( \Upsilon_x, \Upsilon_y )  t}
\textbf{X}_{0},
$
where $\textbf{X}_{0}=\textbf{X}\left(  0\right)  $ represents the initial
condition of state $\textbf{X}$ (i.e., $X^{(0 \sim 3)}$).

The external set of feasibility is given by%
\[
{\footnotesize
\mathbb{F} _{X}^{\mathrm{ex}}=\left\{ \textbf{X} \middle| \begin{array}{c}
	u_2\left( \begin{array}{c}
		\textbf{X},\phi ^{\left( 0\sim 1 \right)},\theta ^{\left( 0\sim 1 \right)},\psi ^{\left( 0\sim 1 \right)},\\
		u_{0}^{\left( 0\sim 2 \right)},u_1,t\\
	\end{array} \right) \in \mathcal{U} _2,\\
	\textbf{X}\in \mathbb{R} ^8,t\ge 0\\
\end{array} \right\}.
}
\]
The overall set of feasibility of subsystem (\ref{eq:UC-FAS_2}) is obtained by
$
\mathbb{F}_{X}=\mathbb{F}_{X}^{\mathrm{in}}\cap\mathbb{F}%
_{X}^{\mathrm{ex}}.
$
Further, we define the overall RoEA of subsystem (\ref{eq:UC-FAS_2}) as%
\[
\mathcal{R}_{X}=\left\{  \textbf{X}_{0}|e^{\mathrm{blkdiag} ( \Upsilon_x, \Upsilon_y )  t}\textbf{X}_{0}\in\mathbb{F}_{X},t\geq0\right\}  .
\]
Then, provided that the initial value $\textbf{X}_{0}$ is selected within
$\mathcal{R}_{X}$, the feasible response $\textbf{X}\left(  t\right)$, i.e., $X^{0 \sim 3}(t)$,
is exponentially convergent within $\mathbb{F}_{X}$.

\subsection{Sub-stabilization synthesis}

As a summary, we define the overall set of feasibility and the overall RoEA by
$\mathbb{F} =\mathbb{F} _z\oplus \mathbb{F} _{\psi}\oplus \mathbb{F} _X$,
$\mathcal{R} =\mathcal{R} _z\oplus \mathcal{R} _{\psi}\oplus \mathcal{R} _X$,
for the UC-FAS model and propose the following theorem.

\begin{theorem}
    The derived 6-DOF quadrotor UC-FAS model \eqref{eq:UC-FAS_0}-\eqref{eq:UC-FAS_2} has a sub-stabilization controller if $\mathcal{R} \neq \emptyset$, and in this case, the controller is given by \eqref{eq:stabilization controller_0}, \eqref{eq:stabilization controller_1}, and \eqref{eq:stabilization controller_2}, with the initial condition $(Z_{0}, \Psi_{0}, \mathbf{X}_{0}) \in \mathcal{R}$. Furthermore, the closed-loop system is composed of linear systems \eqref{eq:closed-loop system_0}, \eqref{eq:closed-loop system_1}, and \eqref{eq:closed-loop system_2}, with responses satisfying $( z^{(0 \sim 1)}(t), \psi^{(0 \sim 1)}(t), X^{(0 \sim 3)}(t) ) \in \mathbb{F}$.
\end{theorem}

Then, the subsequent lemma gives a systematic way to assign the closed-loop systems \eqref{eq:closed-loop system_0}, \eqref{eq:closed-loop system_1}, and \eqref{eq:closed-loop system_2} by a parametric design method.
\begin{lemma}[\cite{Duan_IJSS_VII_Controllability}]   \label{lemma:parametric design approach}
	Let $i \in \{1,2,\ldots,\xi\}$. For any arbitrarily chosen matrix $F_i \in \mathbb{R}^{m_ir_i \times m_ir_i}$, the matrices $A_{i,0\sim m_i-1}$ and $V_i \in \mathbb{R}^{m_ir_i \times m_ir_i}$, which satisfy $\det(V_i) \ne 0$ and the relation
	$
	\Upsilon_i\left( A_{i,0\sim m_i-1}\right)  = V_iF_iV_i^{-1},
	$
	can be determined by
	$
	A_{i,0\sim m_i-1} = -Z_iF_i^{m_i}V_i^{-1}\left(Z_i, F_i \right), 
	$
	where $V_i=V_i\left( Z_i,F_i \right) =[Z_{i}^{T} ~\left( Z_iF_i \right) ^T ~\cdots ~\left( Z_iF_{i}^{m_i-1} \right) ^T]^T,$
	and $Z_i \in \mathbb{R}^{r_i \times m_ir_i}$ is a parameter matrix that must fulfill
	$ \det V_i\left(Z_i, F_i \right) \ne 0.  $
\end{lemma}

The characteristics of closed-loop systems \eqref{eq:closed-loop system_0}, \eqref{eq:closed-loop system_1}, and \eqref{eq:closed-loop system_2}, such as eigenstructure and convergence rate, can be designed by tuning the control gains $A_{0,0\sim 1}$, $A_{1,0\sim 1}$, and $A_{2}$ using the parametric design approach in Lemma \ref{lemma:parametric design approach}.

\section{Tracking control}
Enabling quadrotors to autonomously follow desired trajectories is a common target in many practical applications. This section further explores the tracking control problem using the UC-FAS approach.

Let $z^{\ast}(t)\in \mathbb{R}^{1}$, $\psi^{\ast}(t)\in \mathbb{R}^{1}$, and $X^{\ast}(t)\in \mathbb{R}^{2}$ be reference signals to be tracked by the states $z$, $\psi$, and $X$, respectively, and define the error states as
$
\bar{z}=z-z^{\ast},
\bar{\psi}=\psi -\psi ^{\ast}
$, and
$
\bar{X}=X-X^{\ast},
$
then we have $\bar{z}^{\left( i \right)}=z^{\left( i \right)}-\left( z^{\ast} \right) ^{\left( i \right)},i=0,1,2$, 
$\bar{\psi}^{\left( j \right)}=\psi ^{\left( j \right)}-\left( \psi ^{\ast} \right) ^{\left( j \right)},j=0,1,2$, 
and $\bar{X}^{\left( k \right)}=X^{\left( k \right)}-\left( X^{\ast} \right) ^{\left( k \right)},k=0,1,2,3,4$.
The UC-FAS model \eqref{eq:UC-FAS_0}-\eqref{eq:UC-FAS_2} is converted into the error form
\begin{align}
	&  \ddot{\bar{z}}=u_{0}-g - \ddot{z}^{\ast},  \label{eq:track_UC-FAS_0}\\
	&  \ddot{\bar{\psi}}=u_{1} - \ddot{\psi}^{\ast}, \label{eq:track_UC-FAS_1}\\
	&  \bar{X}^{\left( 4 \right)}=g_{X}( \cdot ) 
		+G_{X}( \cdot ) u_2 - \left( X^{\ast} \right) ^{\left( 4 \right)}.
            \label{eq:track_UC-FAS_2}
\end{align}
The tracking controllers can be similarly designed as
\begin{align}   \label{eq:tracking controller}
	\left\{ \begin{aligned}
		&u_0=-A_{0,0\sim 1}\bar{z}^{\left( 0\sim 1 \right)}+g + \ddot{z}^{\ast},\\
		&u_1=-A_{1,0\sim 1}\bar{\psi}^{\left( 0\sim 1 \right)} + \ddot{\psi}^{\ast},\\
		&u_2=-G_{X}^{-1}\left( \cdot \right) \big( A_{2}\bar{X}^{\left( 0\sim 3 \right)}
		+g_{X}\left( \cdot \right) - \left( X^{\ast} \right) ^{\left( 4 \right)} \big).
	\end{aligned} \right. 
\end{align}
Set of feasibility and RoEA must be similarly analyzed as Section \ref{sec:Sub-stabilization controller design}. Due to page limitations, the analysis is omitted.

\section{Simulation}

We use a simulation to show the feasibility and superiority of the proposed method. The simulated 6-DOF quadrotor model is provided in (\ref{a1})-(\ref{a3}) with the mass $m=0.625\,\mathrm{kg}$, the gravitational acceleration $g=9.8\,\mathrm{m/s}^{2}$, the moment of inertial in the $x$-, $y$-, and $z$-axis $J_{x}=0.0019005\,\mathrm{kg\cdot m}^{2}$, $J_{y}=0.0019536\,\mathrm{kg\cdot m}^{2}$, and $J_{z}=0.0036894\,\mathrm{kg\cdot m}^{2}$.

Using the proposed tracking controller \eqref{eq:tracking controller} and applying the parametric design method outlined in Lemma \ref{lemma:parametric design approach}, we can analytically derive linear constant closed-loop systems with arbitrarily assignable eigenstructures, which are unrealizable through traditional local linearization approaches, nor is it possible with some nonlinear methods like FL. For the subsystems \eqref{eq:track_UC-FAS_0} and \eqref{eq:track_UC-FAS_1} of UC-FAS, the design parameters are specified as $Z_0 = Z_1 = [1 ~1]$ and $F_0 = F_1 = \mathrm{diag}([-4,~-5])$. The corresponding closed-loop systems are obtained as
\begin{align}
	\ddot{\bar{z}}+A_{0,0\sim1}\bar{z}^{\left(  0\sim1\right)  }=0,~
	\ddot{\bar{\psi}}+A_{1,0\sim1}\bar{\psi}^{\left(  0\sim1\right)  }=0,
\end{align}
where the controller gains $A_{0,0\sim1} = A_{1,0\sim1} = [20 ~9]$. For the subsystems \eqref{eq:track_UC-FAS_2} of UC-FAS, note that this subsystem actually consists of two high-order differential equations involving $x$ and $y$. 
Let $Z_{2,x} = Z_{2,y} = [1 ~1 ~1 ~1]$ and $F_{2,x} = F_{2,y} = \mathrm{diag}([-5 ~-6 ~-7 ~-8])$, we obtain the individual controller gains as $A_{2,0\sim 3}^x = A_{2,0\sim 3}^y = [1680 ~1066 ~251 ~26]$, yielding the following closed-loop system:
\begin{align}
	\bar{X}^{\left(  4\right)  }+\mathrm{blkdiag}(A_{2,0\sim 3}^x,A_{2,0\sim 3}^y)\bar{X}^{\left(  0\sim3\right)}=0.
\end{align}

\begin{figure}[htbp]
	\centering
	\includegraphics[scale=0.76]{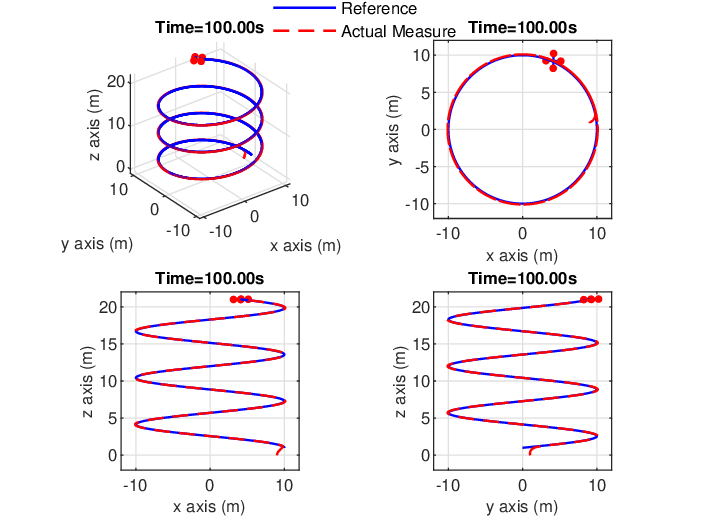}
	\caption{Tracking control follows a 3-D spiral trajectory.}
	\label{fig:fig_3d tracking trajectory}
\end{figure}

\begin{figure}[htbp]
	\centering
	\includegraphics[scale=0.76]{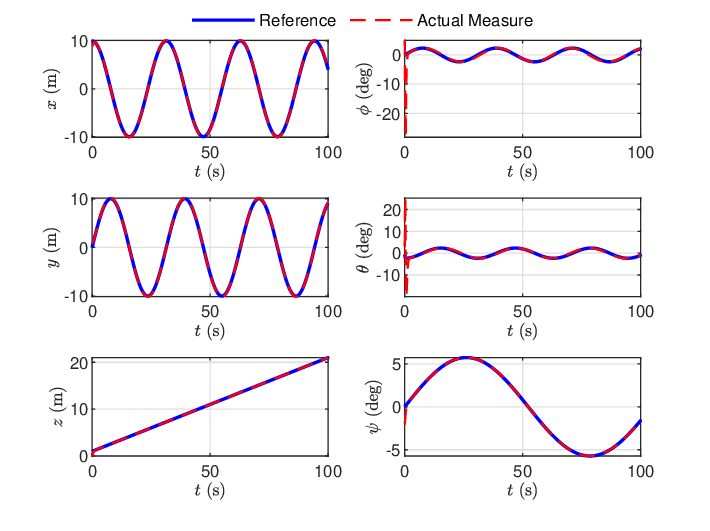}
	\caption{Tracking responses for 6-DOF.}
	\label{fig:fig_tracking response}
\end{figure}

In the simulation, saturation is applied to the quadrotor's actuators to reflect real-world implementation better. The total rotor thrust is limited to the range of $T \in [-100, 100] \,(\mathrm{N})$, and the body torques are restricted to $\tau_{\phi},\tau_{\theta},\tau_{\psi} \in [-0.5, 0.5] \,(\mathrm{N\cdot m})$. 
To validate the control performance of the proposed method, we instruct the quadrotor to follow a spiral trajectory in the 3D space while simultaneously tracking a trigonometric wave for the yaw angle $\psi$ during $0\sim 100\mathrm{s}$. The tracking responses from spatial and temporal perspectives are illustrated in Figs. \ref{fig:fig_3d tracking trajectory} and \ref{fig:fig_tracking response}, respectively. 
The results indicate that the control, based on the proposed UC-FAS approach, achieves excellent performance and demonstrates straightforward design procedures.

\section{Conclusion}

The UC-FAS framework resolves challenges in quadrotor control by removing input-derivative estimation and unifying fragmented FAS modeling efforts. Its ability to decouple dynamics into linear subsystems streamlines controller synthesis and enables performance tuning via eigenstructure assignment, a feature unattainable with most of the existing methods in the state-space framework. 
The simulation results underscore its potential as a versatile tool for underactuated systems, advancing UC-FAS theory toward real-world applicability. Future work will extend this framework to experimental test.

{\footnotesize \bibliography{refs,ref_FAS_Duan,ref_Ren}}

\end{document}